\documentclass[11pt]{llncs}
\usepackage{latexsym,amsfonts,color,amsmath}

\newcommand{\shit}[1]{{\color{black}{#1}}}

\newcommand{\remove}[1]{}

\newcommand{\ex}{{\rm E}}
\newcommand{\e}{{\rm e}}
\newcommand{\eps}{\epsilon}
\newcommand{\p}{\partial}

\newcommand{\N}{{\Bbb N}}

\newcommand{\R}{{\Bbb R}}
\newcommand{\Q}{{\Bbb Q}}

\newcommand{\vv}{{\mathbf v}}
\newcommand{\vx}{{\mathbf x}}
\newcommand{\w}{{\mathbf w}}
\newcommand{\y}{{\mathbf y}}
\newcommand{\z}{{\mathbf z}}
\newcommand{\zzeta}{{\bf \zeta}}

\newcommand{\gmax}{g_{\rm max}}

\newcommand{\proofend}{\medskip \hfill\mbox{$\Box$}\\}
\def\bc{\begin{center}}
\def\ec{\end{center}}

\newcommand{\gnp}{\ensuremath{{\mathcal G}(n,p)}}

\newcommand{\cnd}{\ensuremath{{\mathcal C}_{n,d}}}
\newcommand{\gnd}{\ensuremath{{\mathcal G}_{n,d}}}

\newcommand{\gnm}{\ensuremath{G(n,m)}}

\def\R{\hbox{\rm I\kern-2pt R}}

\def\whp{w.h.p.}
\def\wpp{w.p.p.}

\def\ie{i.e.\ }

\def\3gl{3-{\small{\sc gl}}}

\def\2p{\mbox{$(2+p)$-{\rm SAT}}}

\def\ex{{\bf E}}

\def\algbound3{\tau_3}
\def\algcert3{\sigma_3}

\newcommand{\dima}[1]{{\color{black}{#1}}}
\newcommand{\dimb}[1]{{\color{black}{#1}}}
\newcommand{\cris}[1]{{\color{black}{#1}}}

\title{The Chromatic Number of Random Regular Graphs}

\author{
    Dimitris Achlioptas\inst{1}
    \and
    Cristopher Moore\inst{2}
}

\institute{Microsoft Research, Redmond, WA 98052, USA,\\
\email{optas@microsoft.com}, \and University of New Mexico,
NM 87131, USA,\\\email{moore@cs.unm.edu}}

\begin{document}

\date{}

\maketitle

\begin{abstract}
Given any integer $d \geq 3$, let $k$ be the smallest integer such
that $d < 2k \log k$.  We prove that with high probability the
chromatic number of a random $d$-regular graph {is $k$,
$k+1$, or $k+2$.}
\end{abstract}

\section{Introduction}

In~\cite{luczak2}, {\L}uczak proved that for every real $d>0$
there exists an integer $k=k(d)$ such that \whp\footnote{Given a
sequence of events ${\mathcal E}_n$, we say that ${\mathcal E}$
holds {\em with positive probability} (\wpp) if $\liminf_{n \to
\infty} \Pr[{\mathcal E}_n]
> 0$, and {\em with high probability} (\whp) if $\liminf_{n \to
\infty} \Pr[{\mathcal E}_n] = 1$.}\ $\chi(\mathcal{G}(n,d/n))$ is
either $k$ or $k+1$. Recently, these two possible values were
determined by the first author and Naor~\cite{AchNaor}.

Significantly less is known for random $d$-regular graphs $\gnd$.
In~\cite{friezeluczak},  Frieze and {\L}uczak extended the results
of~\cite{luczak} for $\chi(\gnp)$ to random $d$-regular graphs,
proving that for all integers $d>d_0$, \whp
\[ \left|\chi(\gnd) - \frac{d}{2\log d}\right| =  \Theta\left( \frac{d \,\log \log d}{(\log d)^2} \right) \enspace .
\]

Here we determine $\chi(\gnd)$ up to three possible values for all
integers. Moreover, for roughly half of all integers we determine
$\chi(\gnd)$  up to two possible values. We first replicate the
argument in~\cite{luczak2} to prove
\begin{theorem}\label{thm:twoval}
For every integer $d$, there exists an integer $k=k(d)$ such that
\whp\ the chromatic number of \gnd\ is either $k$ or $k+1$.
\end{theorem}

We then use the second moment method to prove the following.
\begin{theorem}
\label{thm:f} For every integer $d$, \whp\ $\chi(\gnd)$ is either
$k$, $k+1$, or $k+2$, where $k$ is the smallest integer such that
$d < 2k \log k$. If, furthermore, $d > (2k-1) \log k$, then \whp\
$\chi(\gnd)$ is either $k+1$ or $k+2$.
\end{theorem}
The table below gives the possible values of $\chi(\gnd)$ for some
values of $d$.
\[ \begin{array}{c | c c c c c c c}
d    & 4 & 5   & 6 & \mbox{$7,8,9$} & 10 & 100 & \cris{1,\!000,\!000}
\\ \hline \chi(\gnd) & \;\;3,4\;\; & \;\;3,4,5\;\; & \;\;4,5\;\; & \;\;4,5,6\;\; & \;\;5,6\;\; & \;\;18,19,20\;\; &
\;\;46523,46524\;\;
\end{array} \]

\subsection{Preliminaries and outline of the proof}

Rather than proving our results for \gnd\ directly, it will be
convenient to work with random $d$-regular multigraphs, in the
sense of the configuration model~\cite{BoRG}; \cris{that is,} multigraphs
$\cnd$ generated by selecting a uniformly random configuration
(matching) on $dn$ ``vertex copies.'' It is well-known that for any
fixed integer $d$, a random such multigraph is simple \wpp\ As a
result, to prove Theorem~\ref{thm:twoval} we simply establish its
assertion for $\cnd$.

To prove Theorem~\ref{thm:f} we use the second moment method to
show
\begin{theorem}
\label{thm:pos} If $d < 2k \log k$, then \wpp\ $\chi(\cnd) \leq
\cris{k+1}$.
\end{theorem}

\noindent
{\em Proof of Theorem~\ref{thm:f}.} For integer $k$ let $u_k =
(2k-1) \log k$ and $c_k = 2k \log k$. Observe that $c_{k-1} < u_k
< c_k$. Thus, if $k$ is the smallest integer such that $d < c_k$,
then either i) $u_k < d < c_k$ or ii) $u_{k-1} < c_{k-1} < d \leq
u_k < c_k$.

A simple first moment argument (see e.g.\ \cite{molloyms}) implies
that if $d > u_k$ then \whp\ $\chi(\cnd)> k$. Thus, if $u_k < d <
c_k$, then \whp\ $\cnd$ is non-$k$-colorable while \wpp\ it is
$(k+1)$-colorable. Therefore, by Theorem~\ref{thm:twoval}, \whp\
the chromatic number of \cnd\ (and therefore \gnd) is either $k+1$
or $k+2$. In the second case, we cannot eliminate the possibility
that $\gnd$ is \wpp\ $k$-colorable, but we do know that it is
\whp\ non-$(k-1)$-colorable. Thus, similarly, it follows that
$\chi(\gnd)$ is \whp\ $k$, $k+1$ or $k+2$. \proofend

Throughout the rest of the paper, unless we explicitly say
otherwise, we are referring to random multigraphs \cnd. We will
say that a multigraph is $k$-colorable iff the underlying simple
graph is $k$-colorable. Also, we will refer to multigraphs and
configurations interchangeably using whichever form is most
convenient.

\section{2-point concentration}
\label{sec:2point}

\shit{In~\cite{luczak2}, {\L}uczak in fact established two-point
concentration for $\chi(\mathcal{G}(n,d/n))$ for all $\eps>0$ and
$d=O(n^{1/6-\eps})$. Here, mimicking his proof, we establish
two-point concentration for $\chi(\gnd)$ for all $\eps>0$ and
$d=O(n^{1/7-\eps})$.}


Our main technical tool is the following
martingale-based concentration inequality for random variables
defined on \cnd~\cite[Thm 2.19]{wormaldreg}. Given a configuration
$C$, we define a {\em switching\/} in $C$ to be the replacement of
two pairs $\{e_1,e_2\}$, $\{e_3,e_4\}$ by $\{e_1,e_3\}$,
$\{e_2,e_4\}$ or $\{e_1, e_4\}$, $\{e_3,e_2\}$.
\begin{theorem}\label{thm:pairing}
Let $X_n$ be a random variable defined on $\cnd$ such that for any
configurations $C,C'$ that differ by a switching
$$
\left| X_n(C) - X_n(C') \right| \leq b \enspace ,
$$
for some constant $b>0$. Then for every $t>0$,
$$
    \Pr\bigl[X_n \leq \ex[X_n] - t \bigr]
    <
    \e^{-\frac{t^2}{dnb^2}}
    \quad \mbox{and} \quad
    \Pr\bigl[X_n \geq \ex[X_n] + t \bigr]
    <
    \e^{-\frac{t^2}{dnb^2}}
\enspace .
$$
\end{theorem}

Theorem~\ref{thm:twoval} will follow from the following two
lemmata.  The proof of Lemma~\ref{sparseness} is a straightforward
union bound argument and is relegated to the full paper.

\begin{lemma}\label{sparseness}
For any $0 < \eps < 1/6$ and $d < n^{1/6-\eps} $, \whp\ every
subgraph induced by $s \leq nd^{-3(1+2\eps)}$ vertices contains at
most $(3/2-\eps) s$ edges.
\end{lemma}

\begin{lemma}\label{tail}
For a given function $\omega(n)$, let $k=k(\omega,n,p)$ be the smallest $k$ such that
\[ \Pr[\chi(\cnd) \leq k] \geq 1/\omega(n) \enspace . \]
With probability greater than $1 - 1/\omega(n)$, all but $8\sqrt{n
d \log \omega(n)}$ vertices of \cnd\ can be properly colored using
$k$ colors.
\end{lemma}

\noindent 
{\em Proof.} For a \dimb{multigraph} $G$, let $Y_k(G)$ be the
minimal size of a set of vertices $S$ for which $G-S$ is
$k$-colorable. Clearly, for any $k$ and $G$, switching two edges
of $G$ can affect $Y_k(G)$ by at most 4, as a vertex cannot
contribute more than itself to $Y_k(G)$. Thus, if $\mu_k =
\ex[Y_k(\dimb{\cnd})]$, Theorem~\ref{thm:pairing} implies
\begin{equation}\label{pocoloco}
    \Pr[Y_k \leq \mu_k -\lambda \sqrt{n}] < \e^{-\frac{\lambda^2}{16d}}
    \;\; \mbox{{ and }}\;\;
    \Pr[Y_k \geq \mu_k +\lambda \sqrt{n}] < \e^{-\frac{\lambda^2}{16d}}
    \enspace .
\end{equation}
Define now $u=u(n,p,\omega(n))$ to be the least integer for which
$\Pr[\chi(G)\leq u]\geq 1/\omega(n)$. Choosing
$\lambda=\lambda(n)$ so as to satisfy $e^{-\lambda^2/(16 d)}=
1/\omega(n)$, the first inequality in~(\ref{pocoloco}) yields
\[
\Pr[Y_u \leq \mu_u -\lambda \sqrt{n}] < 1/\omega(n)
                                      \leq \Pr[\chi(G)\leq u]
                                      = \Pr[Y_u=0]
                                      \enspace .
\]
Clearly, if $\Pr[Y_u \leq \mu_u -\lambda \sqrt{n}] < \Pr[Y_u=0]$
then  $\mu_u < \lambda \sqrt{n}$. Thus, the second  inequality
in~(\ref{pocoloco}) implies $\Pr[Y \geq 2 \lambda \sqrt{n}]  <
1/\omega(n)$ and, by our choice, $\lambda =4 \sqrt{d
\log\omega(n)}$.
\proofend

\noindent {\bf Proof of Theorem~\ref{thm:twoval}.} \dimb{The
result is trivial for $d=1,2$.} Given $d \geq 3$, let $k= k(d,n)
\geq 3$ be the smallest integer for which the probability that
\dimb{\cnd}\ is $k$-colorable is at least $1/\log\log n$. By
Lemma~\ref{tail}, \whp\ there exists a set of vertices $S$ such
that all vertices outside $S$ can be colored using $k$ colors and
$|S| < 8\sqrt{nd \log \log\log n} <\sqrt{nd} \log n \equiv s_0$.
From $S$, we will construct an increasing sequence of sets of
vertices $\{U_i\}$ as follows. $U_0=S$; for $i\geq 0$, $U_{i+1} =
U_i \cup \{w_1,w_2\}$, where $w_1,w_2 \not\in U_i$ are adjacent
and each of them has some neighbor in $U_i$. The construction
ends, with $U_t$, when no such pair exists.

Observe that the neighborhood of $U_t$ in the rest of the graph,
$N(U_t)$, is always an independent set, \cris{since otherwise the}
construction would have gone on. We further claim that \whp\ the
graph induced by the vertices in $U_t$ is $k$-colorable. Thus,
using an additional color for $N(U_t)$ yields a $(k+1)$-coloring
of the entire \dimb{multigraph}, concluding the proof.

We will prove that $U_t$ is, in fact, 3-colorable by proving that
$|U_{t}| \leq s_0 /\eps$. This suffices since by
Lemma~\ref{sparseness} \whp\ every subgraph $H$ of $b$ or fewer
vertices has average degree less than 3 and hence contains a
vertex $v$ with $\deg(v) \leq 2$. Repeatedly invoking
Lemma~\ref{sparseness} yields an ordering of the vertices in $H$
such that each vertex is adjacent to no more than 2 of its
successors. Thus, we can start with the last vertex in the
ordering and proceed backwards; there will always be at least one
available color for the current vertex. To prove $|U_{t}| \leq
2s_0 \log n$ we observe that each pair of vertices entering $U$
``brings in" with it at least 3 new edges. Therefore, for every $j
\geq 0$, $U_j$ has at most $s_0 + 2j$ vertices and at least $3j$
edges. Thus, by Lemma~\ref{sparseness}, \whp\ $t <
3s_0/(4\eps)$.\proofend


\section{Establishing colorability in two moments}

Let us say that a coloring $\sigma$ is \dimb{{\em
nearly--balanced}} if its color classes differ in size by at most
$1$, and let $X$ be the number of nearly--balanced $k$-colorings of
\dimb{$\cnd$}. Recall that $c_k = 2k \log k$. We will prove that
for all $k \ge 3$ and $d < c_{k-1}$ there exist constants
$C_1,C_2>0$ such that for all sufficiently large $n$ (when $dn$ is
even),
\begin{eqnarray}
    \ex[X]      & > & C_1 \, n^{-(k-1)/2} \, k^n    \left(1-\frac{1}{k}\right)^{dn/2}
    \enspace , \label{eq:down}\\
    \ex[X^2]    & < & C_2 \, n^{-(k-1)}   \, k^{2n} \left(1-\frac{1}{k}\right)^{dn} \enspace
    .\label{eq:up}
\end{eqnarray}
By the Cauchy-Schwartz inequality (see e.g.~\cite[Remark
3.1]{jlr}), we have $\Pr[X>0] > \ex[X]^2/\ex[X^2] \dimb{>
C_1^2/C_2>0}$, and thus Theorem~\ref{thm:pos}.

To prove \eqref{eq:down}, \eqref{eq:up} we will need to bound
certain combinatorial sums up to constant factors. To achieve this
we will use the following Laplace-type lemma, which generalizes a
series of lemmas in~\cite{ksat,hyper,AchNaor}. Its proof is
standard but somewhat tedious, and is relegated to the full paper.

\begin{lemma}
\label{lem:laplace} Let $\ell,m$ be positive integers.  Let $\y
\in \Q^m$, and let $M$ be a $m \times \ell$ matrix of rank $r$
with integer entries whose top row consists entirely of $1$'s.
Let $s,t$ be nonnegative integers, and let $\vv_i, \w_j \in
\N^\ell$ for $1 \le i \le s$ and $1 \le j \le t$, where each
$\vv_i$ and $\w_j$ has at least one nonzero component, and where
moreover $\sum_{i=1}^s \vv_i = \sum_{j=1}^t \w_j$.  Let $f:\R^\ell
\to \R$ be a positive twice-differentiable function.  For $n \in
\N$, define
\[ S_n = \sum_{\{\z \in \N^\ell: M \cdot \z = \y n\}}
\frac{\prod_{i=1}^s (\vv_i \cdot \z)!}{\prod_{j=1}^t (\w_j \cdot
\z)!} \,f(\z/n)^n
\]
and define $g:\R^\ell \to \R$ as
\[ g(\zzeta)
= \frac{\prod_{i=1}^s (\vv_i \cdot \zzeta)^{(\vv_i \cdot \zzeta)}}
   {\prod_{j=1}^t (\w_j \cdot \zzeta)^{(\w_j \cdot \zzeta)}}  \,f(\zzeta)
\]
where $0^0 \equiv 1$.  Now suppose that, conditioned on $M \cdot
\zzeta = \y$, $g$ is maximized at some $\zzeta^*$ with $\zzeta^*_i
> 0$ for all $i$, and write $\gmax = g(\zzeta^*)$.  Furthermore,
suppose that the matrix of second derivatives $g'' = \p^2 g / \p
\zeta_i \,\p \zeta_j$ is nonsingular at $\zzeta^*$.

Then there exist constants $A, B > 0$, such that for any
sufficiently large $n$ for which there exist integer solutions
$\z$ to $M \cdot \z = \y n$, we have
\[ A  \le \frac{S_n}{n^{-(\ell+s-t-r)/2} \,\gmax^n} \le B  \enspace . \]
\end{lemma}

For simplicity, in the proofs of \eqref{eq:down} and \eqref{eq:up}
\cris{below} we will assume that $n$ is a multiple of $k$, so that
nearly--balanced colorings are in fact exactly balanced, with
$n/k$ vertices in each color class. The calculations for other
values of $n$ \cris{differ} by at most a multiplicative constant.

\section{The first moment}
\label{sec:moments}

Clearly, all \dimb{(exactly)} balanced $k$-partitions of the $n$
vertices are equally likely to be proper $k$-colorings. Therefore,
$\ex[X]$ is the number of balanced $k$-partitions, $n! /
(n/k)!^k$, times the probability that a random $d$-regular
configuration is properly colored by a fixed balanced
$k$-partition.

To estimate this probability we will label the $d$ copies of each
vertex, thus giving us $(dn-1)!!$ distinct configurations, and
count the number of such configurations that are properly colored
by a fixed balanced $k$-partition. To generate such a
configuration we first determine the number of edges between each
pair of color classes. Suppose there are $b_{ij}$ edges between
vertices of colors $i$ and $j$ for each $i \ne j$. Then a properly
colored configuration can be generated by i) choosing which
$b_{ij}$ of the $dn/k$ copies in each color class $i$ are matched
with copies in each color class $j \ne i$, and then ii) choosing
one of the $b_{ij}!$ matchings for each unordered pair $i < j$.
Therefore, the total number of properly colored configurations
is
\[ \prod_{i=1}^k \frac{(dn/k)!}{\prod_{j \ne i} b_{ij}!}
\cris{\,\cdot} \prod_{i < j} b_{ij}! = \frac{(dn/k)!^k}{\prod_{i <
j} b_{ij}!} \enspace . \]

Summing over all choices of the $\{b_{ij}\}$ that satisfy the
constraints
\begin{equation}
\label{eq:bij} \forall i: \sum_j b_{ij} = d n / k \enspace ,
\end{equation}
we get
\begin{eqnarray*}
\label{eq:1stmoment}
    \ex[X]  & = &
            \frac{n!}{(n/k)!^k} \frac{1}{(dn-1)!!} \sum_{\{
            b_{ij} \}} \frac{(dn/k)!^k}{\prod_{i < j} b_{ij}!} \\
            & = &
            2^{dn/2} \frac{n!}{(n/k)!^k} \frac{(dn/k)!^k}{(dn)!}
            \sum_{\{ b_{ij} \}} \frac{(dn/2)!}{\prod_{i < j} b_{ij}!}
            \enspace .
\end{eqnarray*}
\remove{Finally, applying Stirling's
approximation~\eqref{eq:stirling}, we get} By Stirling's
approximation $\sqrt{2 \pi n} \,(n/\e)^n \;<\; n! \;<\; \sqrt{4
\pi n} \,(n/\e)^n$ we get
\begin{equation}
\label{eq:1stmoment2} \ex[X] > D_1 \frac{2^{dn/2}}{k^{(d-1)n}}
\sum_{\{ b_{ij} \}} \frac{(dn/2)!}{\prod_{i < j} b_{ij}!} \enspace
,
\end{equation}
where $D_1 = 2^{-(k+1)/2} \,d^{(k-1)/2}$.

To bound the sum in~\eqref{eq:1stmoment2} from below we use
Lemma~\ref{lem:laplace}. {Specifically, $\z$ consists of the
variables $b_{ij}$ with $i < j$, so $\ell=k(k-1)/2$.}  For $k \ge
3$, the $k$ constraints~\eqref{eq:bij} are linearly independent,
so representing them as $M \cdot \z = \y n$ gives a matrix $M$ of
rank $k$.  Moreover, they imply $\sum_{i < j} b_{ij} = dn/2$, so
adding a row of $1$'s to the top of $M$ and setting $y_1 = d/2$
does not increase its rank.  Integer solutions $\z$ exist whenever
$n$ is a multiple of $k$ {and $dn$ is even.}
We set $s=1$ and
$t=\ell$; the vector $\vv_1$ consists of $1$'s and the $\w_j$ are
the $\ell$ basis vectors. Finally, $f(\zzeta)=1$. Thus,
$\ell+s-t-r = -(k-1)$ and
\[ g(\zzeta) = \frac{(d/2)^{d/2}}{\prod_{j=1}^\ell \zeta_k^{\zeta_k}}
= \frac{1}{\prod_{j=1}^\ell (2 \zeta_j / d)^{\zeta_j}} = \e^{(d/2)
H(2 \zzeta/d)} \enspace ,
\]
where $H$ is the entropy function
$H(\vx) = - \sum_{j=1}^\ell x_j \log x_j$.

{Since $g$ is convex it is maximized when $\zzeta^*_j = d/(2\ell)$
for all $1 \le j \le \ell$, and $g''$ is nonsingular. Thus, $\gmax
= (k(k-1)/2)^{d/2}$ implying that for some $A>0$ and all
sufficiently large $n$}
\begin{eqnarray*}
\ex[X]
& > & D_1 \frac{2^{dn/2}}{k^{(d-1)n}}  \times A \,n^{-(k-1)/2} \left(\frac{k(k-1)}{2}\right)^{dn/2} \\
& = & D_1 A \,n^{-(k-1)/2} \,k^n \left(1-\frac{1}{k}\right)^{dn/2}
\enspace .
\end{eqnarray*}
Setting $C_1 = D_1 A$ completes the the proof.

\section{The second moment}

Recall that $X$ is the sum over all balanced $k$-partitions of the
indicators that each partition is a proper coloring. Therefore,
$\ex[X^2]$ is the sum over all pairs of balanced $k$-partitions of
the probability that both partitions properly color a random
$d$-regular configuration. Given a pair of partitions
$\sigma,\tau$, let us say that a vertex $v$ is in class $(i,j)$ if
$\sigma(v)=i$ and $\tau(v)=j$. Also, let $a_{ij}$ denote the
number of vertices in each class $(i,j)$. We call $A=(a_{ij})$ the
{\em overlap matrix} of the pair $\sigma,\tau$.  Note that since
both $\sigma$ and $\tau$ are balanced
\begin{equation}
\label{eq:aconstraints} \forall i: \sum_j a_{ij} = \sum_j a_{ji} =
n/k \enspace .
\end{equation}

We will show that for any fixed pair of $k$-partitions, the
probability that they both properly color a random $d$-regular
configuration depends only on their overlap matrix $A$. Denoting
this probability by $q(A)$, since there are {$n! / \prod_{ij}
a_{ij}!$ pairs of partitions giving rise to $A$,} we have
\begin{equation}
\label{eq:x2} \ex[X^2] = \sum_A \frac{n!}{\prod_{ij} a_{ij}!} \;
q(A)
\end{equation}
where the sum is over matrices $A$
satisfying~\eqref{eq:aconstraints}.

Fixing a pair of partitions $\sigma$ and $\tau$ with overlap
matrix $A$, similarly to the first moment,  we label the $d$
copies of each vertex thus getting $(dn-1)!!$ distinct
configurations. To generate configurations properly colored by
both $\sigma$ and $\tau$ we first determine the number of edges
between each pair of vertex classes. Let us say that there are
$b_{ijk\ell}$ edges connecting vertices in class $(i,j)$ to
vertices in class $(k,\ell)$.  By definition, $b_{ijk\ell} =
b_{k\ell ij}$, and if both colorings are proper, $b_{ijk\ell}=0$
unless $i \ne k$ and $j \ne \ell$. Since the configuration is
$d$-regular, we also have
\begin{equation}
\label{eq:bconstraints}
 \forall i,j:  \sum_{k \ne i, \ell \ne j} b_{ijk\ell} = d a_{ij}  \enspace .
 \end{equation}

To generate a configuration consistent with $A$ and
$\{b_{ijk\ell}\}$ we now i) choose for each class $(i,j)$, which
$b_{ijk\ell}$ of its $d a_{ij}$ copies are to be matched with
copies in each class $(k,\ell)$ with $k \ne i$ and $\ell \ne j$,
and then ii) choose one of the $b_{ijk\ell}!$ matchings for each
unordered pair of classes $i < k$, $j \ne \ell$.  Thus,
\begin{eqnarray}
 q(A) & = & \frac{1}{(dn-1)!!} \sum_{\{b_{ijk\ell}\}}
  \left( \prod_{ij} \frac{(d a_{ij})!}{\prod_{k \ne i, \ell \ne j} b_{ijk\ell}!}
  \cris{\,\,\cdot\!}
  \prod_{i < k, j \ne \ell} b_{ijk\ell}! \right) \nonumber \\
 & = & 2^{dn/2} \,\frac{\prod_{ij} (d a_{ij})!}{(dn)!} \sum_{\{b_{ijk\ell}\}}
  \frac{(dn/2)!}
  {\prod_{i < k, j \ne \ell} b_{ijk\ell}! }
  \label{eq:x2q} \enspace ,
\end{eqnarray}
where the sum is over the $\{b_{ijk\ell}\}$ satisfying~\eqref{eq:bconstraints}.  Combining~\eqref{eq:x2q} with~\eqref{eq:x2} gives
\begin{equation}
\label{eq:x2sum}
 \ex[X^2] = 2^{dn/2} \sum_{\{a_{ij}\}} \sum_{\{b_{ijk\ell}\}}
  \frac{n!}{\prod_{ij} a_{ij}!}
  \,\frac{\prod_{ij} (d a_{ij})!}{(dn)!}
  \frac{(dn/2)!}{\prod_{i < k, j \ne \ell} b_{ijk\ell}! } \enspace
  .
\end{equation}

To bound the sum in~\eqref{eq:x2sum} from above we use
Lemma~\ref{lem:laplace}. We let $\z$ consist of the combined set
of variables $\{a_{ij} \} \cup \{ b_{ijk\ell}: i < k, j \ne
\ell\}$, in which case its dimensionality $\ell$ (not to be
confused with the color $\ell$) is $k^2 + (k(k-1))^2 / 2$.  We
represent the combined system of
constraints~\eqref{eq:aconstraints}, \eqref{eq:bconstraints} as $M
\cdot \z = \y n$.  The $k^2$ constraints~\eqref{eq:bconstraints}
are, clearly, linearly independent while the $2k$
constraints~\eqref{eq:aconstraints} have rank $2k-1$. Together
these imply $\sum_{ij} a_{ij} = 1$ and $\sum_{i < k, j \ne \ell}
b_{ijk\ell} = d/2$, so adding a row of $1$'s to the top of $M$
does not change its rank from $r = k^2 + 2k - 1$.  Integer
solutions $\z$ exist whenever $n$ is a multiple of $k$ and $dn$ is
even.
{Finally, $f(\zzeta) = 2^{d/2}$,} $s = k^2+2$ and $t =
k^2+1+(k(k-1))^2/2$, so $\ell+s-t-r = -2(k-1)$.

Writing $\alpha_{ij}$ and $\beta_{ijk\ell}$ for the components of
$\zzeta$ corresponding to $a_{ij}/n$ and $b_{ijk\ell}/n$,
respectively, we thus have
\begin{eqnarray}
 g(\zzeta) & = & {2^{d/2}}
  \frac{1}{\prod_{ij} \alpha_{ij}^{\alpha_{ij}}}
  \frac{\prod_{ij} (d\alpha_{ij})^{d\alpha_{ij}}}{d^d}
  \frac{(d/2)^{d/2}}{\prod_{i < k, j \ne \ell} \beta_{ijk\ell}^{\beta_{ijk\ell}}} \nonumber \\
& = &
  \frac{1}{\prod_{ij} \alpha_{ij}^{\alpha_{ij}}}
  \frac{d^{d/2} \prod_{ij} \alpha_{ij}^{d\alpha_{ij}}}
     {\prod_{i < k, j \ne \ell} \beta_{ijk\ell}^{\beta_{ijk\ell}}}
     \enspace .
\label{eq:x2g}
\end{eqnarray}

In the next section we maximize $g(\zzeta)$ over $\zzeta \in
\R^{\ell}$ satisfying $M \cdot \zzeta = \y$. We note that $g''$ is
nonsingular \cris{at the maximizer we find below, but we relegate
the proof of this fact to the full paper.}

\section{A tight relaxation}

Maximizing $g(\zzeta)$ over $\zzeta \in \R^{\ell}$ satisfying $M
\cdot \zzeta = \y$ is greatly complicated by the constraints
\begin{equation}
\forall i,j: \sum_{k \ne i, \ell \ne j} \beta_{ijk\ell} = d
\alpha_{ij} \label{eq:betaconstraints} \enspace .
\end{equation}
To overcome this issue we i) reformulate $g(\zzeta)$ and ii) relax
the constraints, in a manner such that the maximum value remains
unchanged while \cris{the} optimization becomes much easier.

The relaxation amounts to replacing the $k^2$
constraints~\eqref{eq:betaconstraints} with their sum divided by
2, i.e., with the single constraint
\begin{equation}
 \sum_{i < k, j \ne \ell} \beta_{ijk\ell} = d/2 \enspace .
 \label{eq:single}
\end{equation}
But attempting to maximize \eqref{eq:x2g} under this single
constraint is, in fact, a bad idea since the new maximum is much
greater. Instead, we \dimb{maximize the following equivalent form
of $g(\zzeta)$}
\begin{equation}
 \label{eq:rewriteg}
g(\zzeta) = \frac{1}{\prod_{ij} \alpha_{ij}^{\alpha_{ij}}}
  \,\frac{d^{d/2} \prod_{ij} \alpha_{ij}^{\sum_{k \ne i, \ell \ne j} \beta_{ijk\ell}}}
     {\prod_{i < k, j \ne \ell} \beta_{ijk\ell}^{\beta_{ijk\ell}}}
     \enspace ,
\end{equation}
\dimb{derived by} using~\eqref{eq:betaconstraints} to substitute
for the exponents $d \alpha_{ij}$ in the numerator of
\eqref{eq:x2g}. This turns out to be enough to drive the maximizer
back to the subspace $M \cdot \zzeta = \y$.

Specifically, let us hold $\{ \alpha_{ij} \}$ fixed and maximize
$g(\zzeta)$ with respect to $\{ \beta_{ijk\ell} \}$ using the
method of Lagrange multipliers.  Since $\log g$ is monotonically
increasing in $g$, it is convenient to maximize $\log g$ instead.
If $\lambda$ is the Lagrange multiplier corresponding to the
constraint~\eqref{eq:single}, we have {for all $i < k, j \ne
\ell$:}
\begin{eqnarray*}
 \lambda & = & \frac{\p}{\p \beta_{ijk\ell}} \log g(\zzeta)
 = \frac{\p}{\p \beta_{ijk\ell}}
    \left( \beta_{ijk\ell} \log (\alpha_{ij} \alpha_{k\ell}) - \beta_{ijk\ell} \log \beta_{ijk\ell} \right) \\
 & = & \log \alpha_{ij} + \log \alpha_{k\ell} - \log \beta_{ijk\ell} - 1
\end{eqnarray*}
and so
\begin{equation}
 \forall i < k, j \ne l: \;
 \beta_{ijk\ell} = C \alpha_{ij} \alpha_{k\ell},
 \mbox{ where } C = \e^{-\lambda-1} \enspace .
\label{eq:lagrange}
\end{equation}
{Clearly, such $\beta_{ijk\ell}$ also satisfy the original
constraints~\eqref{eq:betaconstraints},
\cris{and therefore the upper bound we obtain from this relaxation
is in fact tight.}

To solve for $C$ we sum~\eqref{eq:lagrange} and
use~\eqref{eq:single}, getting
\[ \frac{2}{C} \sum_{i < k, j \ne \ell} \beta_{ijk\ell} = \frac{d}{C}
= \sum_{i \ne k, j \ne \ell} \alpha_{ij} \alpha_{k\ell} = 1 -
\frac{2}{k} + \sum_{ij} \alpha_{ij}^2 \equiv p \enspace . \] Thus
$C=d/p$ and~\eqref{eq:lagrange} becomes
\begin{equation}
\forall i < k, j \ne l: \;
\beta_{ijk\ell} =
\frac{d \alpha_{ij} \alpha_{k\ell}}{p}
\label{eq:b}
\end{equation}

Observe that $p = p(\{a_{ij}\})$ is the probability that a single
edge whose endpoints are chosen uniformly at random is properly
colored by both $\sigma$ and $\tau$, if the overlap matrix is
$a_{ij} = \alpha_{ij} n$. Moreover, the values for the
$b_{ijk\ell}$ are exactly what we would obtain, in expectation, if
we chose from among the ${n \choose 2}$ edges with replacement,
rejecting those improperly colored by $\sigma$ or $\tau$, until we
had $dn/2$ edges---in other words, if our graph model was $G(n,m)$
with replacement, rather than $\gnd$.

Substituting the values~\eqref{eq:b} in~\eqref{eq:rewriteg} and
applying~\eqref{eq:single} yields the following upper bound on
$g(\zzeta)$:
\begin{eqnarray*}
g(\zzeta) & \le &
\frac{1}{\prod_{ij} \alpha_{ij}^{\alpha_{ij}}}
\,\frac{d^{d/2} \prod_{ij} \alpha_{ij}^{(d/p) \alpha_{ij} \sum_{i \ne k, j \ne \ell} \alpha_{k\ell}}}
   {(d/p)^{\sum_{i < k, j \ne \ell} \beta_{ijk\ell}}
      \prod_{i < k, j \ne \ell} (\alpha_{ij} \alpha_{k\ell})^{(d/p) \alpha_{ij} \alpha_{k\ell}} }  \\
& = & \frac{1}{\prod_{ij} \alpha_{ij}^{\alpha_{ij}}}
\,\frac{d^{d/2}}{(d/p)^{d/2}} \left (\frac{\prod_{ij} a_{ij}^{\alpha_{ij} \sum_{i \ne k, j \ne \ell} \alpha_{k\ell}}}
   {\prod_{i \ne k, j \ne \ell} \alpha_{ij}^{\alpha_{ij} \alpha_{k\ell}}} \right)^{\!d/p} \\
& = &
\frac{p^{d/2} }{\prod_{ij} \alpha_{ij}^{\alpha_{ij}}}\\
& \equiv & g_{\gnm}(\{ \alpha_{ij} \}) \enspace .
\end{eqnarray*}


In~\cite[Thm 5]{AchNaor}, Achlioptas and Naor showed that for $d <
c_{k-1}$ the function $g_{\gnm}$ is {maximized when $\alpha_{ij} =
1/k^2$ for all $i,j$.}  In this case $p = (1-1/k)^2$, implying
\[ \gmax \le k^2 p^{d/2} = k^2 \left( 1 - \frac{1}{k} \right)^d \]
and, therefore, that for some constant $C_2$ and sufficiently
large $n$
\[ \ex[X^2] \le {C_2} \,n^{-(k-1)} \,k^{2n} \left(1 - \frac{1}{k} \right)^{dn} \enspace
.\]
%
%

\section{Directions for further work}

\paragraph{A sharp threshold for regular graphs.} It has long
been conjectured that for every $k>2$, there exists a critical
constant $c_k$ such that a random graph $G(n,m=cn)$ is \whp\
$k$-colorable if $c< c_k$ and \whp\ non-$k$-colorable if $c>c_k$.
It is reasonable to conjecture that the same is true for random
regular graphs, \ie that for all $k>2$, there exists a critical
integer $d_k$ such that a random graph $\gnd$ is \whp\
$k$-colorable if $d\leq d_k$ and \whp\ non-$k$-colorable if
$d>d_k$. If this is true, our results imply that for $d$ in
``good'' intervals $(u_k, c_k)$ \whp\ the chromatic number of
$\gnd$ is precisely \cris{$k+1$}, while for $d$ in ``bad''
intervals $(c_{k-1},u_k)$ the chromatic number is \whp\ either $k$
or $k+1$.

\paragraph{Improving the second moment bound.} Our proof
establishes that if $X,Y$ are the numbers of balanced
$k$-colorings of $\gnd$ and $G(n,m=dn/2)$, respectively, then
$\ex[X]^2 / \ex[X^2] = \Theta(\ex[Y]^2/\ex[Y^2])$. Therefore, any
improvement on the upper bound for $\ex[Y^2]$ given
in~\cite{AchNaor} would immediately {give}
an improved positive-probability $k$-colorability result for $\gnd$.

In particular, Moore has conjectured that the function $g_{\gnm}$
is maximized by matrices with a certain form. If true, this
immediately gives an improved lower bound, $c^*_k$, for
$k$-colorability satisfying $c^*_{k-1} \to u_k - 1$. 
{This} would shrink the union of the ``bad" intervals to a set of
measure 0, with each such interval containing precisely one
integer $d$ for each $k\geq k_0$.

\paragraph{3-colorability of random regular graphs.}
It is {easy to show} that a random 6-regular graph is \whp\
non-3-colorable. \dimb{On the other hand, in~\cite{4col} the
authors showed that 4-regular graphs are \wpp\ 3-colorable.} Based
on considerations from statistical physics,
Krz\c{a}ka{\l}a, Pagnani and Weigt~\cite{weigt} have
conjectured that a random $5$-regular graph is \whp\
$3$-colorable. The authors (unpublished) have shown that applying
the second moment method to the number of balanced 3-colorings
cannot establish this fact (even with positive probability).

\bigskip \noindent
{\bf Acknowledgments.} C. Moore is grateful to Tracy Conrad, Alex
Russell, and Martin Weigt for helpful conversations,  and is
supported by NSF grant PHY-0200909.

\end{document}